\documentclass[prc,preprint,preprintnumbers,amsmath,amssymb]{revtex4-1}

\usepackage{bm,curves}
\usepackage{epsfig}
\usepackage[colorlinks=true,linkcolor=blue]{hyperref} 
\usepackage{color}
\usepackage{dcolumn}
\usepackage{amsmath}
\usepackage{natbib}

\DeclareGraphicsExtensions{.ps}

\newcommand{\lan}{\langle}
\newcommand{\ran}{\rangle}
\newcommand{\EA}{{\it et al.}}

\def\gtorder{\mathrel{\raise.3ex\hbox{$>$}\mkern-14mu
        \lower0.6ex\hbox{$\sim$}}}
\def\ltorder{\mathrel{\raise.3ex\hbox{$<$}\mkern-14mu
        \lower0.6ex\hbox{$\sim$}}}

\newcommand{\nc}{\newcommand}   
\nc{\mb}[1]{\makebox[#1]{}}
\nc{\V}{{\rm v}}
\nc{\W}{{\scriptscriptstyle W}}
\nc{\X}{{\scriptscriptstyle X}}
\nc{\CSV}{{\scriptscriptstyle CSV}}
\nc{\ra}{\rightarrow}
\nc{\alS}{{\alpha_{\scriptscriptstyle S}}}
\nc{\aSpi}{{\frac{\alS}{2\pi}}}
\nc{\api}{{\frac{\alpha}{2\pi}}}
\nc{\dwtilm}{{\delta \widetilde{m}}}
\nc{\ppg}{\pi^+\pi^-\gamma}
\nc{\nubar}{{\overline{\nu}}}
\nc{\nuN}{{\nu N_0}}
\nc{\nubN}{{\overline{\nu} N_0}}
\nc{\snuNC}{{\langle \sigma^{\nuN}_{\NC}\rangle }}
\nc{\snubNC}{{\langle \sigma^{\nubN}_{\NC}\rangle }}
\nc{\snuCC}{{\langle \sigma^{\nuN}_{\CC}\rangle }}
\nc{\snubCC}{{\langle \sigma^{\nubN}_{\CC}\rangle }}
\nc{\snNC}{{\langle \sigma^{\nu p}_{\NC}\rangle }}
\nc{\snbNC}{{\langle \sigma^{\nubar p}_{\NC}\rangle }}
\nc{\snCC}{{\langle \sigma^{\nu p}_{\CC}\rangle }}
\nc{\snbCC}{{\langle \sigma^{\nubar p}_{\CC}\rangle }}
\nc{\Rnu}{{R^{\nu}}}
\nc{\Rnub}{{R^{\overline{\nu}}}}
\nc{\sintW}{{\sin^2 \theta_{\W} }}
\nc{\vp}{{\bf p}}
\nc{\uv}{{u_{\rm v}}}
\nc{\dv}{{d_{\rm v}}}
\nc{\ubar}{{\overline{u}}}
\nc{\dbar}{{\overline{d}}}
\nc{\sbar}{{\overline{s}}}
\nc{\cbar}{{\overline{c}}}
\nc{\Ubar}{{\overline{U}}}
\nc{\Dbar}{{\overline{D}}}
\nc{\Sbar}{{\overline{S}}}
\nc{\Qbar}{{\overline{Q}}}
\nc{\FbWp}{{\overline{F}_2^{Wp}}}
\nc{\FbWD}{{\overline{F}_2^{WD}}}
\nc{\rz}{{1\over \rho_0^2}}
\nc{\gLu} {{g_L^u}}
\nc{\gRu} {{g_R^u}}
\nc{\gLd} {{g_L^d}}
\nc{\gRd} {{g_R^d}}
\nc{\Delu} {{\Delta u^2}}
\nc{\Deld} {{\Delta d^2}}
\nc{\Rnp} {{R^{\nu}_p}}
\nc{\Rnbp} {{R^{\nubar}_p}}
\nc{\Pcs}{{P_{CS}}}
\def\CC{{\scriptscriptstyle CC}}

\def\NC{{\scriptscriptstyle NC}}

\nc{\be}{\begin{equation}}
\nc{\ee}{\end{equation}}
\nc{\bea}{\begin{eqnarray}}
\nc{\eea}{\end{eqnarray}}
\nc{\F}{{\scriptscriptstyle F}}  
\nc{\xF}{{x_{\F}}}
\nc{\Fcc}{F_2^c}
\nc{\rmE}{{\rm E}}

\def\IE{{\it i.e.,}}
\def\EG{{\it e.g.,}}
\def\EA{{\it et al.}}

\def\psl{p\hspace*{-0.17cm}\slash\hspace*{0.022cm}}
\def\ksl{\displaystyle{\not} k }
\def\lsl{l\hspace*{-0.17cm}\slash\hspace*{0.022cm}}

\def\lsim{\mathrel{\rlap{\lower4pt\hbox{\hskip1pt$\sim$}}
    \raise1pt\hbox{$<$}}}
\def\gsim{\mathrel{\rlap{\lower4pt\hbox{\hskip1pt$\sim$}}
    \raise1pt\hbox{$>$}}}

\def \beqn{\begin{eqnarray}}
\def \eeqn{\end{eqnarray}}

\def \bea{\begin{eqnarray}}
\def \beq{\begin{equation}}
\def \eea{\end{eqnarray}}
\def \eeq{\end{equation}}

\def \bwt{\begin{widetext}}
\def \ewt{\end{widetext}}
\def \lan{\langle }
\def \ran{\rangle }
\def \ti{\widetilde}

\graphicspath{{./WF-Fig/}}

\begin{document}

\preprint{NT@UW-16-08}

\title{A Euclidean bridge to the relativistic constituent quark model}

\author{T. J. Hobbs$^1$}
\email{tjhobbs@uw.edu}

\author{Mary Alberg$^{1,2}$, Gerald A. Miller$^1$}
\affiliation{
        $^1$\mbox{Department of Physics,
         University of Washington, Seattle, Washington 98195, USA} \\
        $^2$Department of Physics,
         Seattle University,
         Seattle, Washington 98122, USA
}

\date{\today}
\begin{abstract}
\begin{description}
\item[Background]
Knowledge of nucleon structure is today ever more of a precision science,
with heightened theoretical and experimental activity expected in coming years.
At the same time, a persistent gap lingers between theoretical approaches
grounded in Euclidean methods (\EG~lattice QCD, Dyson-Schwinger Equations [DSEs])
as opposed to traditional Minkowski field theories (such as light-front constituent
quark models).
\item[Purpose]
Seeking to bridge these complementary worldviews, we explore the potential of a {\it Euclidean
constituent quark model} (ECQM). This formalism enables us to study the gluonic dressing of the
quark-level axial-vector vertex, which we undertake as a test of the framework.
\item[Method]
To access its indispensable elements with a minimum of inessential detail, we develop
our ECQM using the simplified quark\! $+$\! scalar diquark picture of the nucleon. We
construct a hyperspherical formalism involving polynomial expansions of diquark
propagators to marry our ECQM with the results of Bethe-Salpeter Equation (BSE) analyses,
and constrain model parameters by fitting electromagnetic form factor data.
\item[Results]
From this formalism, we define and compute a new quantity --- the {\it Euclidean density function}
(EDF) --- an object that characterizes the nucleon's various charge distributions as functions of
the quark's Euclidean momentum. Applying this technology and incorporating information from
BSE analyses, we find the dressing effect on the proton's axial-singlet charge
to be small in magnitude and consistent with zero.

\item[Conclusions]
The scalar quark\! $+$\! diquark ECQM is a step toward a realistic quark model in
Euclidean space, and urges additional refinements. The small size we obtain for the impact
of the dressed vertex on the axial-singlet charge suggests that models without
this effect are on firm ground to neglect it.
\end{description}
\end{abstract}
\maketitle

%
\section{Introduction}
\label{sec:intro}
%
%
Hadronic physics is presently at an important crossroads. On the one hand, with its
advantageous representation of Minkowski field theory, light-front formalism
\cite{Chang:1973qi,Lepage:1980fj,Miller:2000kv,Brodsky:2000ii,Bakker:2013cea,Brodsky:1997de}
has made impressive gains in understanding the proton's flavor and spin structure
\cite{Cardarelli:1995dc,Carbonell:1998rj,Miller:2002ig,Cloet:2012cy}. At much the same
time, techniques grounded in Euclidean field theory, such as Lattice QCD \cite{Aoki:2016frl}
and the methodology of Bethe-Salpeter Equations (BSEs)
\cite{Salpeter:1951sz,Nambu:1961tp,Maris:2003vk,Roberts:1994dr,Cloet:2013jya},
continue to unfold an ever more refined picture of the hadronic spectrum, as well as its
various excitations and transitions. An effort to reconcile these two families of
approaches is therefore more of a crying necessity than ever before. The present analysis
represents an initial step to {\bf bridge} this enduring gap by formulating a {\it Euclidean
constituent quark model} (ECQM).

To this end, we craft a simple model in Euclidean space which binds the constituent
quark into the nucleon through the exchange of a scalar spectator diquark. While the
quark-diquark approach itself is hardly new (such models have an established history
in the analyses of both the DIS sector \cite{Meyer:1990fr,Mulders:1992za,Jakob:1997wg,Gross:2012sj}
and elastic scattering \cite{Cloet:2012cy}), our specific formulation of a Euclidean
constituent quark model has not to our knowledge been previously attempted.

Standard light-front theory \cite{Tiburzi:2002sw,Miller:2009fc} extracts bound state properties (\EG~elastic
form factors, inelastic structure functions) from overlaps of $3$-dimensional light-front wave functions (LFWFs),
which are themselves obtained by integrating a $4$-dimensional Bethe-Salpeter amplitude over the ``minus''
components of the internal momenta $k^- \equiv k^0 - k^3$; these in turn provide
a means of relating the constituent quark model to form factors and GPDs
\cite{Diehl:2003ny,Belitsky:2005qn,GonzalezHernandez:2012jv,Hobbs:2014lea}.
Despite the remarkable success of methods rooted in constituent quark models, an uncircuitous means of
relating them to Euclidean approaches remains lacking, however. That is, although techniques for projecting,
\EG~the pion's Bethe-Salpeter amplitude onto the LF have been pioneered recently \cite{Chang:2013pq},
a direct formulation of the quark model in Euclidean space has not yet been put forth.
The chief aim of the present article is to do precisely this, leading to the aforementioned
ECQM. However, the implementation in Euclidean space requires techniques inspired by hyperspherical
QED calculations \cite{Blum:2002ii,Levine:1974xh,Roskies:1990ki,Rosner:1967zz}, which we trace in detail
in Sec.~\ref{sec:HypS} below.
Following angular integration of the resulting $4$-dimensional amplitudes in Euclidean hyperspherical space,
the formalism we develop outputs distributions for the quark-level densities of the proton as functions of
the intermediate quark's Euclidean momentum. These latter quantities we designate {\it Euclidean density
functions} (EDFs), and we carry out their evaluation in the sections below.

In the present paper, we test our formalism by performing an analysis of the
quark helicity share of the proton's spin by evaluating the flavor-singlet axial charge
as spelled out in later sections. The origin of the proton's spin in the angular
momentum of its QCD constituents is a problem that has bedeviled hadronic physics ever
since the advent of the ``spin crisis'' in the late 1980s following the revelation
\cite{Ashman:1987hv,Ashman:1989ig} of the European Muon Collaboration (EMC) concerning
the small size of the proton's integrated spin-dependent structure function,
$\int_0^1 g^p_1(x)\, dx = 0.114 \pm 0.012 \pm 0.026$. During the intervening decades,
sufficient progress has been made to reduce the crisis to a mere ``spin problem'' as it
is now more commonly known. Even so, the exact interplay of the various relevant dynamics
\cite{Myhrer:1988ap,Myhrer:2007cf,Thomas:2008ga} remains sufficiently subtle as to prevent
an unambiguous reckoning of the multiple effects giving rise to the proton's spin.

Canonically, the spin of the proton is decomposed among contributions from quark and gluon
helicity and orbital angular momentum as \cite{Jaffe:1995an,Ji:1996ek,Filippone:2001ux}
\begin{equation} 
{1 \over 2}\ =\ {1 \over 2}\, \sum_q \Delta q\ +\ L_q\ +\ J_g\ ,
\label{eq:spin}
\end{equation} 
and the contribution from the total quark helicity $\sum_q \Delta q$ is now understood to represent
approximately one third of the total nucleon spin, and has been the focus of intense experimental and 
theoretical effort \cite{Bluemlein:2002be,Leader:2005ci,Hirai:2006sr,deFlorian:2005mw,Chambers:2015bka}.
Despite recent progress \cite{Cloet:2012cy}, obtaining this result in the context of constituent quark
models, including those formulated on the light-front, remains an elusive goal.
For this reason, an assessment of the r\^ole played by the exchange of nonperturbative gluons in the
setting of a constituent quark model could help weigh whether this effect substantially alters the spin decomposition
of Eq.~(\ref{eq:spin}). To accomplish this, we use the aforementioned hyperspherical ECQM to incorporate
information from BSE analyses on the quark's dressed axial-vector vertex
\cite{Maris:1997hd,Bhagwat:2002tx,Bhagwat:2007ha,Chang:2012cc}, ultimately finding a minimal effect.
The remainder of the paper is organized as follows: Sec.~\ref{sec:EM} treats the standard covariant approach, with a description of the formalism
needed to fit current data in the elastic electromagnetic sector with the bare ECQM in Sec.~\ref{sec:EM-A}, and
a prediction of the proton's axial-singlet charge in Sec.~\ref{sec:a0c}; Sec.~\ref{sec:HypS} describes the hyperspherical formalism. Herein, the basic
properties of EDFs are introduced in Sec.~\ref{sec:EDF}, and the simplest nontrivial calculation --- the EDF for the proton's charge distribution
--- is given in Sec.~\ref{sec:F1}. Having thus completely determined the details of the bare hyperspherical ECQM, we use it to predict the
axial-singlet charge of the proton in Sec.~\ref{sec:a0}, as well as the distribution of this axial charge as a function of the struck quark
Euclidean momentum $\ti{k}$. In Sec.~\ref{sec:gluon} we fold the latest numerical estimates for the soft gluon dressing effect on the axial
charge of an individual quark into our formalism, and draw our final conclusions in Sec.~\ref{sec:conc}. Lastly, select formulae are postponed
 to Appendices \ref{sec:appA} and \ref{sec:appB}. 
\begin{figure}
\includegraphics[scale=0.45]{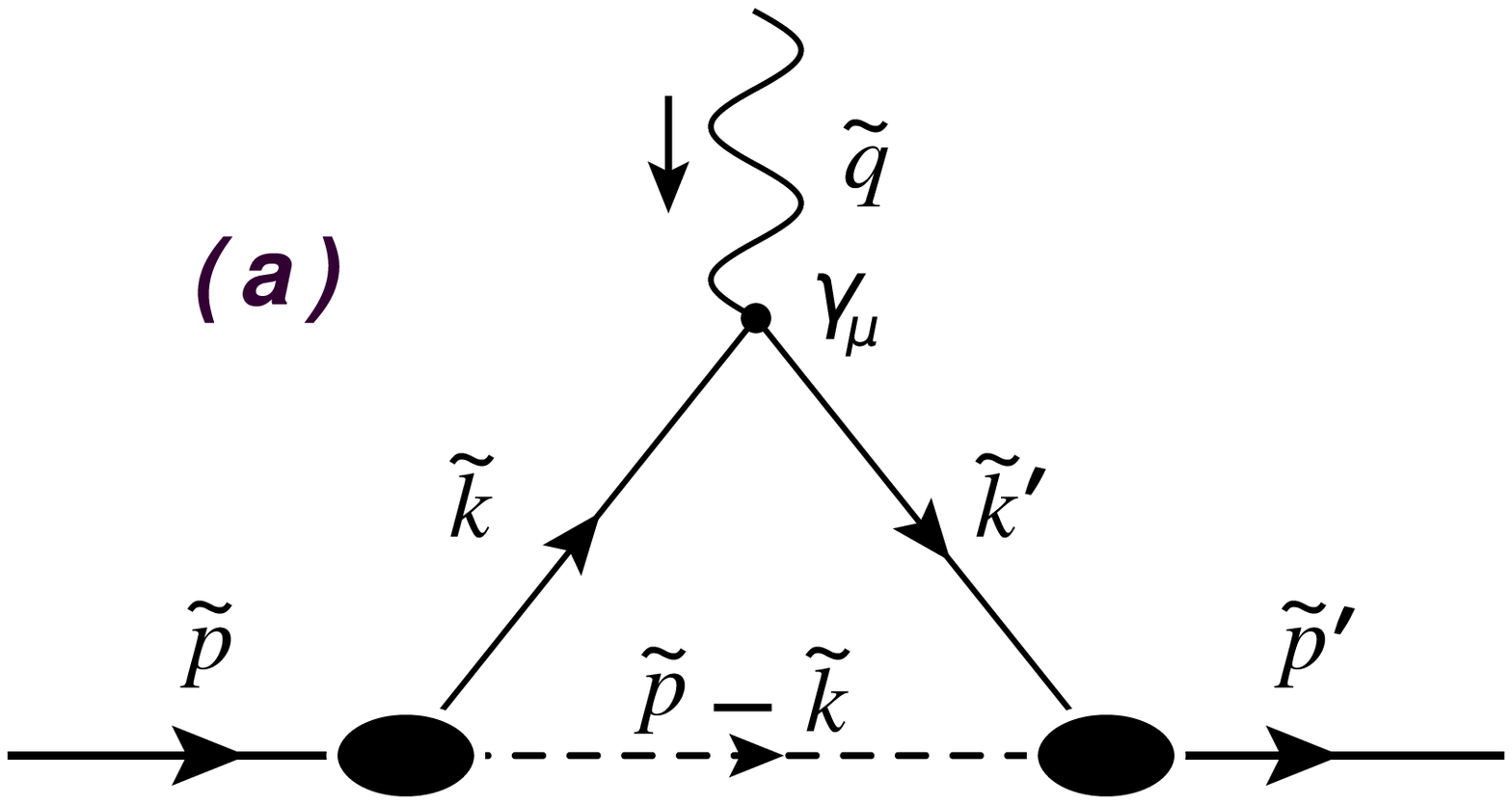} \ \ \ \ \ \ \
\includegraphics[scale=0.45]{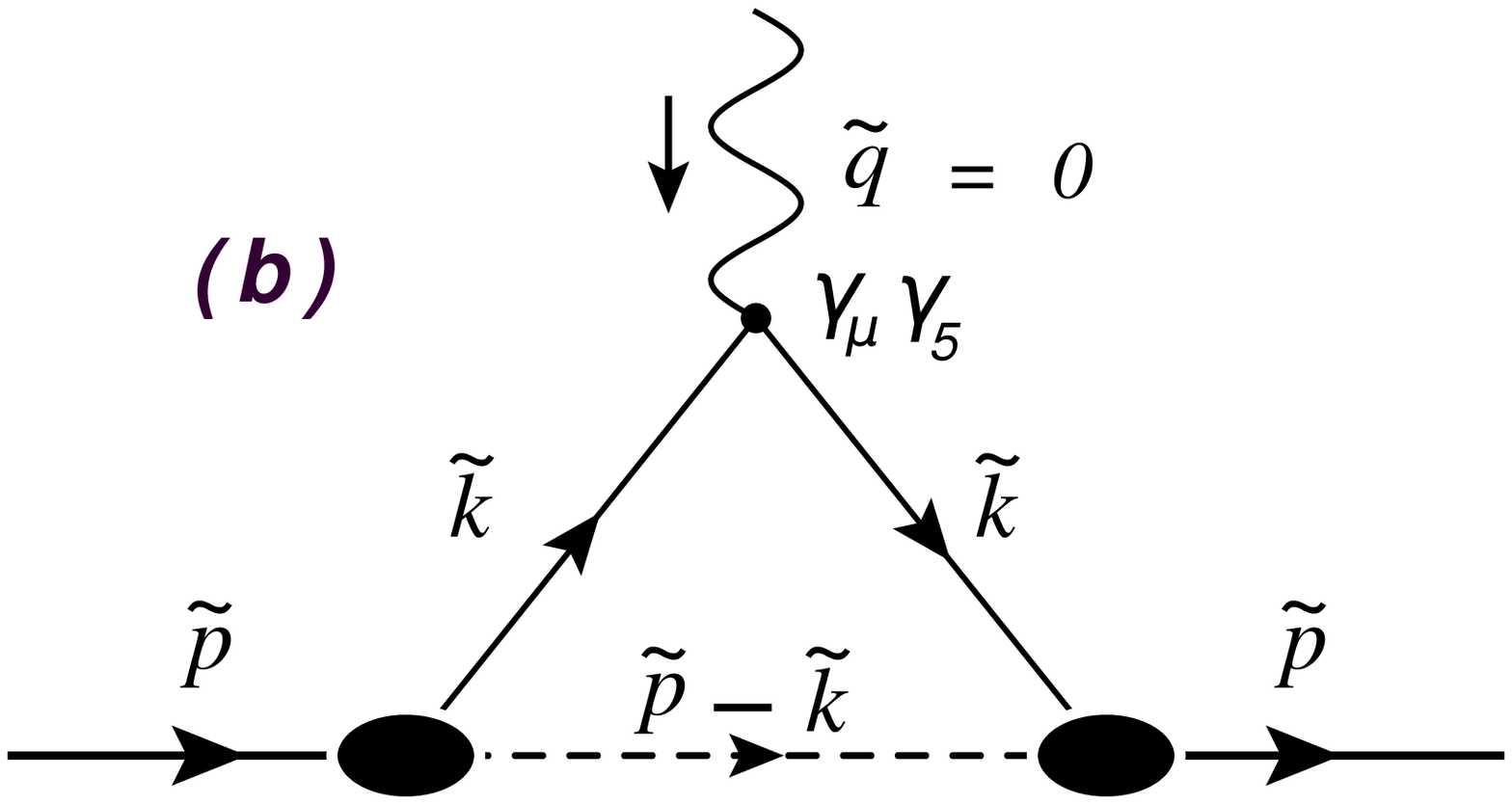}
\caption{ (Color online)
({\bf a}) The standard diagram responsible for the first nontrivial contribution to $F_{1,2}(\ti{q}^2)$.
({\bf b}) The main graph for the quark contributions to the nucleon's axial-singlet charge $a_0$. In both
cases, solid internal lines represent the propagation of the interacting quark, while the dashed lines are
for the scalar spectator diquark. The ovate blobs symbolize our prescription for the momentum dependence of
the nucleon-quark-diquark interaction as given by $\varphi(\ti{k}^2)$ in Eq.~(\ref{eq:vertex}).
}
\label{fig:FF}
\end{figure}
%
\section{The bare model: electromagnetic structure and spin }
\label{sec:EM}
\subsection{Electromagnetic form factors}
\label{sec:EM-A}
In the quark\! $+$\! scalar diquark picture,
computing the Pauli and Dirac form factors $F_1(\ti{q}\!\;^2)$ and $F_2(\ti{q}\!\;^2)$ as functions of the spacelike
photon virtuality squared $\ti{q}\!\;^2$ amounts to evaluating the leading triangle diagram in Fig.~\ref{fig:FF}({\bf a}),
which here represents an amplitude formulated in Euclidean space. For this purpose, we take the propagators of the
scalar diquark (of mass $m_D$) and quark (of mass $m$) to be, respectively,
\begin{align}
D\big( [\ti{p} - \ti{k}]^2 \big)\ &=\ { 1 \over [\ti{p} - \ti{k}]^2 + m_D^2 }\ , \nonumber \\
S\big(\!\;\ti{k}\!\;\big)\ &=\ { 1 \over i\ti{\ksl} + m }\ ,
\end{align}
where we in general denote Euclidean $4$-vectors as $\ti{v}_\mu$, and the main prescription-dependent ingredient
of the ECQM involves making a formal choice to characterize the binding of the struck constituent quark into
the nucleon. To accomplish this, it is necessary to stipulate a relativistic vertex factor for the momentum
dependence of the nucleon-quark-diquark interaction, represented by the ``blobs'' appearing in both panels of
Fig.~\ref{fig:FF}. The systematics involved in the implementation of such phenomenological vertex factors have
been explored in diverse contexts, including in models of nucleon structure
\cite{Zoller:1991cb,Melnitchouk:1993nk,Melnitchouk:1996fj} and nuclear scattering \cite{Bodek:1980ar}; in the
end, however, we select for simplicity a minimal choice consistent with Lorentz covariance: a scalar function
of the quark's Euclidean $4$-momentum $\ti{k}$ with the general form
\begin{equation}
\varphi(\ti{k}^2)\ \equiv\ g\ \left( { \Lambda^2 \over \ti{k}^2 + \Lambda^2 } \right)\ .
\label{eq:vertex}
\end{equation}
Of course other analytic forms for the vertex function may also be used ({\it e.g.,}
multipoles involving higher powers, or functions of the spectator diquark $4$-momentum),
but these ultimately lead to qualitatively similar results, and in practice we find use of
Eq.~(\ref{eq:vertex}) simplifies calculations dramatically. For this reason, the remainder
of the present analysis is carried out using Eq.~(\ref{eq:vertex}). In light of our choice
for the nucleon-quark-diquark vertex function, the model parameters in our framework are
thus the strength of the nucleon's couplings to its internal quark/diquark degrees of
freedom $g$ (which acts as an overall normalization), the constituent masses of the
quark and scalar diquark $m$ and $m_D$, respectively, and the ultraviolet cutoff
parameter $\Lambda$, all of which we take from fits in the electromagnetic sector. Namely,
the form factors $F_{1,2}(\ti{q}^2)$ are extracted from the triangle diagram shown
in Fig.~\ref{fig:FF}, which gives the extended electromagnetic vertex
$\Gamma_\mu \big(\ti{p}^{\,\prime},\ti{p}\!\;\big)$ of the nucleon as
\begin{align}
\label{eq:EM-amp}
\overline{u}(\ti{p}^{\,\prime})\, \gamma_\mu\, u(\ti{p}\!\;)\ &\longrightarrow\ \overline{u}(\ti{p}^{\,\prime})\,
\Gamma_\mu\big(\ti{p}^{\,\prime},\ti{p}\!\;\big)\, u(\ti{p}\!\;) \\
&=\ {1 \over (2\pi)^4}\ \int d^4 \ti{k}\ \overline{u}(\ti{p}^{\,\prime})
\left( {1 \over i\ti{\ksl}^{\,\prime} + m } \right)\, \gamma_\mu\,
\left( {1 \over i\ti{\ksl} + m } \right) u(\ti{p}\!\;)
\left( {\varphi(\ti{k}^{\,\prime 2})\, \varphi(\ti{k}^2) \over [\ti{p}-\ti{k}]^2 + m^2_D } \right)\ , \nonumber
\end{align}
where $\ti{p}$ $\big( \ti{p}^{\,\prime} \big)$ is the intial (final) proton $4$-momentum, $\ti{k}\!\;^\prime = \ti{k} + \ti{q}$,
and $\ti{p}^{\,\prime} = \ti{p} + \ti{q}$. Using the general form of the photon-proton vertex given by
Eq.~(\ref{eq:EMV}) in App.~\ref{sec:appA}, we compute this latter amplitude using standard techniques
\cite{Peskin:1995ev,Miller:2009sg} involving Feynman parameters and momentum shifts to obtain
\begin{align}
\label{eq:F1cov}
F_1(\ti{q}\!\;^2)\ &=\ \left({ g \Lambda^2 \over 4\pi }\right)^2
\int_0^1 dx \int_0^{1-x} dy \int_0^{1-x-y} dz \int_0^{1-x-y-z} dw \\
& \hspace*{6.3cm} \times\ \left( \left[{1 \over \Delta^2}\right]^2 
+ 2 N_1(\ti{q}\!\;^2)\ \left[{1 \over \Delta^2} \right]^3 \right)\ ,\nonumber\\
F_2(\ti{q}\!\;^2)\ &=\ 2 \left({ g \Lambda^2 \over 4\pi }\right)^2
\int_0^1 dx \int_0^{1-x} dy \int_0^{1-x-y} dz \int_0^{1-x-y-z} dw\
N_2(z)\ \left[{1 \over \Delta^2} \right]^3\ ,
\label{eq:F2cov}
\end{align}
in which
\begin{align}
N_1(\ti{q}\!\;^2)\ &=\ \big(m + zM\big)^2 - (x+w)\,\big(1-x-z-w\big)\, \ti{q}\!\;^2\ , \\
N_2(z)\ &=\ 2M\, (1-z)\, \big(m + zM\big)\ , \\
\Delta^2\ &=\ (x+w)\,\big(1-x-z-w\big)\, \ti{q}\!\;^2 + (x+y)\, m^2 + zm^2_D \nonumber\\
&\ \hspace*{2.7cm} - z\,(1-z)M^2 + (1-x-y-z)\, \Lambda^2\ .
\end{align}
Above, $M$ is the mass of the on-shell nucleon, and we have made use of the Euclidean Gordon Identity
given by Eq.~(\ref{eq:Gordon}) to decompose the amplitude of Eq.~(\ref{eq:EM-amp}) into separate Pauli
and Dirac components {\it \`a la} Eq.~(\ref{eq:EMV}).

With these explicit expressions for $F_1$ and $F_2$, it is simple
to construct the familiar Sach's parametrization of the nucleon's electromagnetic form factors:
\begin{align}
G_E(\ti{q}\!\;^2)\ &\equiv\ F_1(\ti{q}\!\;^2)\ -\ {\ti{q}\!\;^2 \over 4 M^2}\ F_2(\ti{q}\!\;^2)\ , \nonumber\\
G_M(\ti{q}\!\;^2)\ &\equiv\ F_1(\ti{q}\!\;^2)\ +\ F_2(\ti{q}\!\;^2)\ ,
\label{eq:sachs}
\end{align}
\begin{figure}
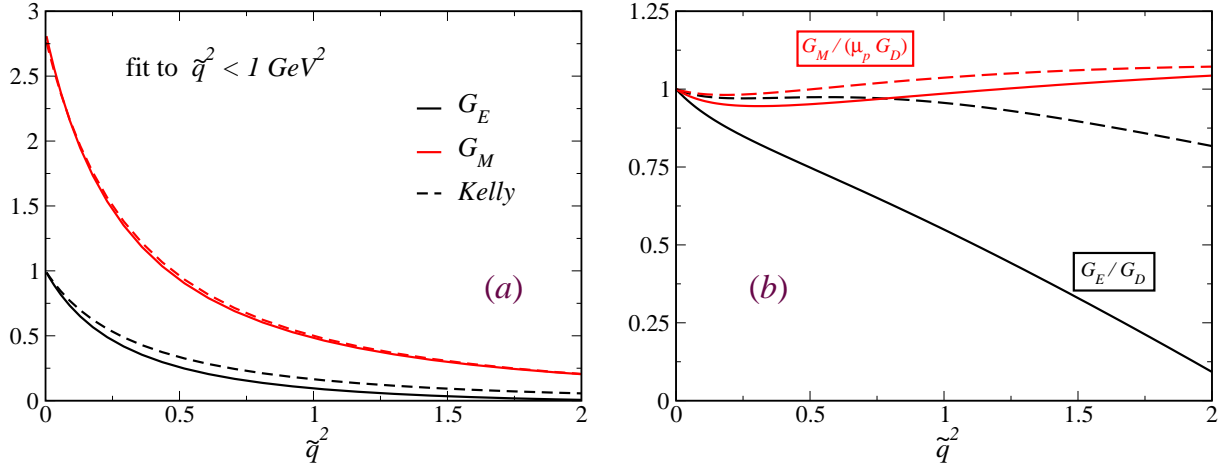

\includegraphics[scale=0.3]{F2a.eps} \ \ \
\includegraphics[scale=0.3]{F2b.eps}
\caption{ (Color online)
({\bf a}) A plot of the fitted electromagnetic form factors $G_{E,M}(\ti{q}^2)$, where we constrain fits
with the phenomenological parametrization of Kelly \cite{Kelly:2004hm} for $\ti{q} \le 1$ GeV. Here,
solid lines give the result of our fitted model for the parameters listed in
Table~\ref{tab:fit}, while the dashed lines are the phenomenological fits of Ref.~\cite{Kelly:2004hm},
with $G_E$ given in black and $G_M$ in red in both cases. ({\bf b}) A similar comparison, but in this
case for the form factor ratios with respect to the well-known dipole parametrization \cite{Kelly:2004hm}
$G_D(\ti{q}^2) \equiv (1 + \ti{q}^2 \big/ \Lambda^2_D)^{-2}$, where $\Lambda^2_D=0.71$ GeV$^2$.
}
\label{fig:EM_fits}
\end{figure}
\hspace{-0.31cm} and we may determine the model parameters by fitting these expressions to experimental
data on the proton.  For this purpose, we treat the phenomenological parametrization of Kelly \cite{Kelly:2004hm}
as a proxy for the world's experimental data and global fits thereof \cite{Arrington:2006zm,Perdrisat:2006hj},
rather than preferencing individual sets; we may then minimize the numerical badness-of-fit measure
\begin{equation}
\chi^2\ \equiv\
{1 \over 2 n_p}\ \sum_{i=1}^{n_p} \left( {G_E(\ti{q}\!\;^2_i) - G^{\mathit phen.}_E(\ti{q}\!\;^2_i) \over G^{\mathit phen.}_E(\ti{q}\!\;^2_i)} \right)^2\
+\ \left( {G_M(\ti{q}\!\;^2_i) - G^{\mathit phen.}_M(\ti{q}\!\;^2_i) \over G^{\mathit phen.}_M(\ti{q}\!\;^2_i)} \right)^2\ .
\label{eq:chi2}
\end{equation}
We note that to ensure the numerical validity of the hyperspherical Euclidean space formalism presented later in Sec.~\ref{sec:HypS},
we in practice find it necessary to constrain the value of the diquark mass to be no less than that of the proton,
$m_D \ge M$, while the other parameters are allowed to float freely over a broad range. This condition is a generic artifact
of hyperspherical techniques as applied to massive theories \cite{Roskies:1990ki}, and for QED can be circumvented
with an appropriate deformation of the integration contour in the complex $\ti{k}^2$ plane. For the amplitudes under consideration
here, however, such an approach meets further complications due to the presence of quark denominators $\sim\! (\ti{k}^2 + m^2)^{-2}$,
which can produce singularities in the timelike region $\ti{k}^2 < 0$ into which the contour over $\ti{k}^2$ is deformed; we therefore
opt for the simpler $m_D \ge M$ condition in this initial study. We note of course that this procedure confers the added benefit of
simulating the effects of a confining potential in the sense that the nucleon is thereby prohibited from decaying into its constituents
($m+m_D \ge M$, for any choice of $m$).
Also, for the sake of describing the nucleon axial-singlet charge (which is defined at $\ti{q}\!\;^2=0$) we concentrate our fits at low
photon virtualities, and hence only constrain them with experimental information for $\ti{q} \le 1$ GeV. Doing so,
we find that fitting our scalar diquark model to the Kelly prediction for $G_E$ and $G_M$ at 5 uniformly-chosen points in the
domain $0 \le \ti{q} \le 1$ GeV [\IE~$n_p=5$ in Eq.~(\ref{eq:chi2}) above] results in the description plotted in
Fig.~\ref{fig:EM_fits}, which corresponds to a $\chi^2$ per datum of $0.003$ for the specific parameter values given
in Table~\ref{tab:fit}. The numerical values of the fitting parameters imply a mass for the diquark comparable to
that of the nucleon (consistent with Faddeev Equation studies, \EG~Ref.~\cite{Segovia:2015ufa}), and a rather large constituent
quark mass $m \sim 600$ MeV.
\begin{table*}[tbp]
\begin{center}
\addtolength{\tabcolsep}{5.0pt}
\addtolength{\extrarowheight}{1.8pt}
\begin{tabular*}{\textwidth}{c@{\extracolsep{\fill}}|ccc|ccccc}
\hline\hline
$\chi^2$           & $m$     &  $m_D$    &   $\Lambda$    & $g$     &  $\mu_p~(\mu_N)$ &  $a_0$   &  $M^1_{\overline{f}_1}$  &  $M^1_{\overline{a}_0}$\\
\hline
0.00297 \ \ \ \    & 0.637   &  0.947    &     0.228      & 79.104  &        2.843     &   0.784  &         0.1985           &        0.08125         \\
\hline\hline
\end{tabular*}
\caption{
The collection of parameters that follow from constraining our model to the proton electromagnetic form factors 
$G_E$ and $G_M$ at low $\ti{q} \le 1$ GeV as given by \cite{Kelly:2004hm}. The parameters given in
the first enclosed box are fitted directly, while those in the open box at the far right are predicted by the
fitted model. Note that the interaction strength $g$ and bare axial-singlet charge $a_0$ determined in
Sec.~\ref{sec:EM} are dimensionless, while the final two columns give the first moments of the electric and
axial-singlet quark charge EDFs $M^1_{\overline{f}_1}$ and $M^1_{\overline{a}_0}$ in GeV$^2$; units
elsewhere are in GeV unless otherwise noted.
}
\label{tab:fit}
\end{center}
\end{table*}
In particular, the two panels of Fig~\ref{fig:EM_fits} compare this fitted model to the parametrization of
Ref.~\cite{Kelly:2004hm} for the proton, both at the level of the separate form factors $G_E$ and $G_M$ themselves
({\bf a}), as well as for the instructive ratios ({\bf b}) with respect to the one-parameter dipole approximation
\cite{Kelly:2004hm} $G_D(\ti{q}\!\;^2) \equiv (1 + \ti{q}\!\;^2 \big/ \Lambda^2_D)^{-2}$, with $\Lambda^2_D=0.71$ GeV$^2$
--- the latter serving to draw attention to subtleties in the form factors' behavior at larger $\ti{q}\!\;^2$. In both panels also,
solid curves represent the output of our fitted model, while dashed lines are the prediction of Ref.~\cite{Kelly:2004hm}.
For the region of interest ($\ti{q}\!\;^2 \gtrsim 0$), fitted results agree especially well with $G_M$, matching
its qualitative dependence on $\ti{q}\!\;^2$ quite closely; for $G_E$, however, the agreement is somewhat weaker,
as especially highlighted by the relatively steep downturn of the solid-black curve of Fig.~\ref{fig:EM_fits}({\bf b}).
At the same time, we adjudicate the better-than $\sim\! 10\%$ agreement at lowest $\ti{q}\!\;^2 \lesssim 0.2$ GeV$^2$
for $G_E$ and percent-level agreement for $G_M$ to be fully adequate for our demonstration of the hyperspherical
formalism here, which we pursue in the following sections only for quantities defined in the real limit, $\ti{q}\!\;^2 = 0$,
including the axial charge $a_0$.
%
%
\subsection{Axial-singlet charge}
\label{sec:a0c}
The total quark helicity contribution to the nucleon spin in Eq.~(\ref{eq:spin}) may be identified with
the matrix element for the axial-singlet charge of the proton \cite{Li:2015exr}, $a_0 = \sum_q \Delta q$,
which we write explicitly as
\begin{align}
2M \ti{S}_\mu\, a_0\ \equiv\ \langle \ti{p}, s |\ \overline{q}\, \gamma_\mu\! \gamma_5\, q\ | \ti{p}, s \rangle\ ,
\hspace*{1.0cm}
\ti{S}_\mu\ \equiv\ \frac{1}{2M}\ \overline{u}(\ti{p}\!\;)\, \gamma_\mu\! \gamma_5\, u(\ti{p}\!\;)\ ,
\label{eq:a0-def}
\end{align}
in which $\ti{S}_\mu$ represents the nucleon's Euclidean spin $4$-vector, which obeys $\ti{S} \cdot \ti{p} = 0$ and
$\ti{S}^2 = -1$.
For the non-pointlike proton basis states consistent with the bare quark\! $+$\! diquark picture, the matrix element of
Eq.~(\ref{eq:a0-def}) can be realized diagrammatically in a triangle graph akin to that which produced
Eqs.~(\ref{eq:F1cov}) and (\ref{eq:F2cov}) for the proton's electromagnetic substructure --- albeit with the appropriate
$\sim\! \gamma_\mu\! \gamma_5$ operator entering at the axial current-quark vertex. This is shown explicitly in
Fig.~\ref{fig:FF}({\bf b}), wherein $\ti{p}\!\;^\prime = \ti{p}$, as is relevant for the axial-singlet charge defined at
$\ti{q} = 0$. Using our established Euclidean conventions, this then gives the amplitude
\begin{align}
2M \ti{S}_\mu\, a_0\ =\ {1 \over (2\pi)^4}\ \int d^4 \ti{k}\ \overline{u}(\ti{p}\!\;)
\left( {1 \over i\ti{\ksl} + m } \right)\ \gamma_\mu\! \gamma_5\
\left( {1 \over i\ti{\ksl} + m } \right) u(\ti{p}\!\;)
\left( {|\varphi(\ti{k}^2)|^2 \over [\ti{p}-\ti{k}]^2 + m^2_D } \right)\ .
\label{eq:a0_amp}
\end{align}
Thus we can follow a procedure similar to that used in the electromagnetic sector to
compute the bare (\IE~{\it undressed}) quark\! $+$\! scalar diquark model prediction
for the proton's axial-singlet charge, keeping in mind that we will ultimately match
our ECQM formalism to the standard calculation in Sec.~\ref{sec:a0}, constituting
a vital test. We find
\begin{align}
\label{eq:a0_cov1}
2M \ti{S}_\mu a_0\ &=\ \Gamma(5)\, {g^2 \Lambda^4  \over (2\pi)^4}\ \int\, {d^4 \ti{l} \over \large( \ti{l}^2 + \Delta^2 \large)^5}\
\int dx\, dy\, dz\,\ xy\ \delta\big(1 - [x+y+z] \big) \nonumber\\
&\ \ \ \ \ \ \times\ \overline{u}(\ti{p}\!\;)
\left( { -i(\ti{\lsl} + z\ti{\psl}) + m } \right)\, \gamma_\mu\! \gamma_5\,
\left( { -i(\ti{\lsl} + z\ti{\psl}) + m } \right) u(\ti{p}\!\;)\ ;
\end{align}
again using textbook \cite{Peskin:1995ev} covariant methods, this can be manipulated to yield
\begin{align}
\label{eq:a0_cov2}
a_0\ &=\ -\left({g \Lambda^2 \over 4\pi}\right)^2\ \int_0^1 dy \int_0^{1-y} dz\ 
y (1-y-z) \left( \left[{1 \over \Delta^2} \right]^2 - 2(m+zM)^2 \left[{1 \over \Delta^2}\right]^3 \right)\ ,
\end{align}
where here the explicit expression for the denominator in terms of masses and Feynman parameters is
\begin{equation}
\Delta^2\ =\ (1-y-z)m^2 + y\Lambda^2 + z\!\:m^2_D -z(1-z)M^2\ ,
\end{equation}
and we have implemented the shift $\ti{k}_\mu \rightarrow \ti{l}_\mu = \ti{k}_\mu - z\,\ti{p}_\mu$,
and made use of Eq.~(\ref{eq:EuDir}). Thus, Eq.~(\ref{eq:a0_cov2}) is fully defined, and may be computed
with the model parameters determined in the electromagnetic sector --- \IE~the values contained within
the inner box of Table~\ref{tab:fit}. Inserting these, we get $a_0 = 0.784$, which we also report in the rightmost
partition of Table~\ref{tab:fit}. We reproduce this value via hyperspherical techniques in Sec.~\ref{sec:a0}.
%
%
\section{Hyperspherical formalism}
\label{sec:HypS}
\subsection{Euclidean density function}
\label{sec:EDF}
Here we introduce the framework necessary to obtain $4$-dimensional Euclidean quark-level
densities --- for the proton's electromagnetic charge in Sec.~\ref{sec:F1}, and its
axial-singlet charge in Sec.~\ref{sec:a0}.

Formally, we seek $4$-dimensional densities dependent on the interacting quark's Euclidean
momentum $\ti{k}$. Such quantities would be analogous to the squares of Bethe-Salpeter
wave functions $\Psi(k;p)$ from which LFWFs can be derived via the appropriate integral over
$\int dk^-$ at fixed LF time \cite{Tiburzi:2002sw,Miller:2009fc} as described in
Sec.~\ref{sec:intro}.
%
%
Properly formulated, in our case these density functions will allow the recovery of bulk properties of the
nucleon from radial integrals in Euclidean space governed by the parameters of a constituent
quark model. That is, the total nucleon charge and axial-singlet charge follow from the zeroth
moment of the {\it Euclidean density functions} (EDFs) $\overline{f}_1(\ti{k}^2)$ and
$\overline{a}_0(\ti{k}^2)$, respectively:
\begin{align}
\label{eq:F1mom}
F_1(\ti{q}\!\;^2\!=\!0)\ &=\ \int d\ti{k}^2\ \overline{f}_1(\ti{k}^2)\ , \\
a_0\ &=\ \int d\ti{k}^2\ \overline{a}_0(\ti{k}^2)\ ,
\label{eq:a0mom}
\end{align}
where the integrations over $\int\! d\ti{k}^2$ remain after summing over angles, and
EDFs for other charges may also be constructed.
In fact, inasmuch as EDFs enjoy the proper support (in this case, vanishing in the
limit $\ti{k}^2 \to \infty$), their lowest moments in $\ti{k}^2$ may also be computed:
\begin{align}
\label{eq:f1k2_def}
M^n_{\bar{f}_1}\ &\equiv\ \int d\ti{k}^2\ \left( \ti{k}^2 \right)^n\, \overline{f}_1(\ti{k}^2)\ , \\
M^n_{\bar{a}_0}\ &\equiv\ \int d\ti{k}^2\ \left( \ti{k}^2 \right)^n\, \overline{a}_0(\ti{k}^2)\ ,
\label{eq:a0k2_def}
\end{align}
for which the choice $(n=0)$ corresponds to the expressions in Eqs.~(\ref{eq:F1mom}) and (\ref{eq:a0mom}),
while the nontrivial first moments $(n=1)$, corresponding to $M^1 \sim \lan \ti{k}^2 \ran$, provide information
on the mean $\ti{k}^2$ of the electromagnetic and axial-charge densities. We determine these explicitly
in Secs.~\ref{sec:F1} and \ref{sec:a0} below, and ultimately plot their associated integrands in
Fig.~\ref{fig:F1a0}.

Pending this more detailed calculation, the proton's charge EDF $\overline{f}_1(\ti{k}^2)$ may be described to
first approximation in the spirit of Feynman~\EA~\cite{Feynman:1971wr}, using a Euclideanized Gaussian wave
function $\psi(\ti{k}^2) \sim \exp(-R^2 \ti{k}^2 \big/ 2)$:
\begin{align}
F_1(\ti{q}\!\;^2\!=\!0)\ &=\ {1 \over (2\pi)^4}\, \int d^4\ti{k}\ \big| \psi(\ti{k}^2) \big|^2\ =\ 1  \nonumber\\
&\rightarrow\ \psi(\ti{k}^2)\ =\ \left( 4\pi R^2 \right)\, \exp \left\{ -{1 \over 2}\,R^2\, \ti{k}^2 \right\}\ ,
\label{eq:WF}
\end{align}
for which the dependence of the wave function on the quark momentum $\ti{k}$ is governed purely by the
proton RMS radius, $R \equiv \lan r^2_p \ran^{1/2} \approx 0.88\, \mathrm{fm} = 1\big/(0.227\, \mathrm{GeV})$ \cite{Mohr:2015ccw}.
Noting Eq.~(\ref{eq:radint}), we conclude
\begin{equation}
\overline{f}^{\mathrm{WF}}_1(\ti{k}^2)\ =\ R^4\, \ti{k}^2\ \exp \left\{ -R^2\, \ti{k}^2 \right\}\ ,
\label{eq:f1WF}
\end{equation}
a simple result to which we compare the model results of Secs.~\ref{sec:F1} and \ref{sec:a0} below as an
instructive benchmark. Plotting the integrand $2\ti{k}\,\overline{f}^{\mathrm{WF}}_1(\ti{k}^2)$ of $F_1(0)$ against
$\ti{k}$ in Fig.~\ref{fig:F1a0}, the resulting distribution peaks predictably near $\ti{k} \gtrsim 0.2$ GeV due
to our numerical choice of $R$, but then has a sharper momentum dependence at higher $\ti{k}$ not found for
the more realistic model calculations presented below; this fact alone highlights the necessity for the more detailed
hyperspherical treatment of nucleon spin structure outlined in Secs.~\ref{sec:F1}--\ref{sec:a0}.
Ultimately, in a utilitarian sense the EDFs of Eqs.~(\ref{eq:F1mom}) and (\ref{eq:a0mom}) also permit
an interface with the output of traditional Euclidean field-theoretic approaches, as emphasized in
Sec.~\ref{sec:intro}. Whereas the formalism of Sec.~\ref{sec:EM} is adequate for the
determination of the total proton charge and helicity in the bare quark model, we ultimately wish to absorb
the results of BSE analyses into our ECQM to assess the gluon dressing effect. For this
purpose, however, BSEs describe the impact of soft gluon exchange in the form of vertex functions of the
quark's Euclidean momentum, and there is no straightforward way to incorporate such quantities into the
bare calculation of Sec.~\ref{sec:a0c}, especially given the reliance of the latter upon shifting loop
momenta away from those given in Fig.~\ref{fig:FF}({\bf b}).

On the other hand, given their status as vertex functions of the quark momentum, BSE results may be incorporated
directly into the integrated EDFs typified by 
Eq.~(\ref{eq:a0mom}) as quark momentum-dependent smearing functions $f_g(\ti{k}^2)$. It is precisely such
a scheme that we pursue here for the quark helicity contribution to the nucleon spin, $a_0$.
Thus, with the EDF $\overline{a}_0(\ti{k}^2)$ and the smearing function $f_g(\ti{k}^2)$ for the gluon-dressing
effect in hand, one may compute the impact of soft gluon exchange upon the total quark helicity
contribution to the proton spin, leading to a corrected axial-singlet charge
\begin{equation}
a'_0\ =\ \int d\ti{k}^2\ \overline{a}_0(\ti{k}^2)\ f_g(\ti{k}^2)\ ,
\label{eq:aprime}
\end{equation}
where in practice we identify the gluonic smearing function with the nonperturbative axial-vector
vertex factor of BSE studies, $f_g(\ti{k}^2) = F_R(\ti{k}^2,0)$, which we take from Ref.~\cite{Chang:2012cc}
and describe in greater detail in Sec.~\ref{sec:gluon}. Moreover, we point out that assuming the perturbative
result expected to hold at $\ti{k} \gg 0$ for the gluon dressing function, $f_g(\ti{k}^2) = 1$, in Eq.~(\ref{eq:aprime})
simply recovers the bare ECQM calculation given by Eq.~(\ref{eq:a0mom}).
We can in fact achieve the specifics of the general formalism described above, and this amounts to the main
result of the present paper. We derive the EDFs of Eqs.~(\ref{eq:F1mom}) and (\ref{eq:a0mom}) by closely
following the analogous calculation for the hadronic vacuum polarization effect in the muon's anomalous
magnetic moment \cite{Blum:2002ii}; viz., we now evaluate Eq.~(\ref{eq:EM-amp}) for $\ti{p}\!\;^\prime = \ti{p}$
in Sec.~\ref{sec:F1} and Eq.~(\ref{eq:a0_amp}) in Sec.~\ref{sec:a0} using a hyperspherical formalism
originally adapted to QED \cite{Levine:1974xh,Roskies:1990ki,Rosner:1967zz}.
%
%
\subsection{Quark charge distribution}
\label{sec:F1}
The hyperspherical formalism we describe below is of sufficient generality that it may be deployed in the evaluation
of various Euclidean momentum distributions. As an initial demonstration, however, we highlight the calculation of the
EDF for the proton's electric charge: \IE~the integrand leading to $F_1(\ti{q}\!\;^2\!=\!0)$ of
Eq.~(\ref{eq:F1mom}). As will be the case for the subsequent determination of $\overline{a}_0(\ti{k})$, we start at
amplitude-level, in this case with Eq.~(\ref{eq:EM-amp}), which at $\ti{q}\!\;^2=0$ yields
\begin{align}
2i\,\ti{p}_\mu\, F_1(0)\ &=\ {1 \over (2\pi)^4}\ \int d^4 \tilde{k}\ \overline{u}(\ti{p}\!\;)
\left( {1 \over i\ti{\ksl} + m } \right)\, \gamma_\mu\,
\left( {1 \over i\ti{\ksl} + m } \right) u(\ti{p}\!\;)
\left( { \big|\varphi(\ti{k}^2)\big|^2  \over [\ti{p}-\ti{k}]^2 + m^2_D } \right)\ \\
&=\ {g^2 \Lambda^4 \over (2\pi)^4}\ \int d\ti{k}^4\
{\overline{u}(\ti{p}\!\;)
\left[ -2\ti{\ksl}\,\ti{k}_\mu + \left( \ti{k}^2 + m^2 \right) \gamma_\mu - im \{\gamma_\mu , \ti{\ksl} \} \right]
u(\ti{p}\!\;)
\over
(\ti{k}^2+m^2)^2\, (\ti{k}^2+\Lambda^2)^2\, \left( [ \ti{p}-\ti{k} ]^2 + m^2_D \right)\ 
}\ ,
\label{eq:F1noCon}
\end{align}
where we have again used Eq.~(\ref{eq:EMV}) for the general form of the electromagnetic vertex given in App.~\ref{sec:appA}.
To apply the hyperspherical formalism, we must express the numerator algebra leading to $F_1(0)$ in terms
of inner products. For this example, we achieve this by contracting both sides of Eq.~(\ref{eq:F1noCon})
with $\ti{p}_\mu$ and using the identities of App.~\ref{sec:appA}, which brings us to the expression
\begin{align}
F_1(0)\ &=\ {g^2 \Lambda^4 \over (2\pi)^4}\ \int d\ti{k}^4\
{
\ti{k}^2 + m^2 -{2 \over \ti{p}\!\;^2}\, \big(\ti{p} \cdot \ti{k} \big)^2 + {2 \over \ti{p}\!\;^2} m M \big( \ti{p}\cdot\ti{k} \big)
\over
(\ti{k}^2+m^2)^2\, (\ti{k}^2+\Lambda^2)^2\, \left( [ \ti{p}-\ti{k} ]^2 + m^2_D \right)
}\ .
\label{eq:F1inner}
\end{align}
More critically, rather than shifting away the term in the denominator
$\sim\!\! (\ti{p} \cdot \ti{k})$ as in the standard covariant calculations involving Feynman parameters
[Eqs.~(\ref{eq:F1cov}) -- (\ref{eq:F2cov}) and (\ref{eq:a0_cov2})], we instead make an expansion of
the scalar diquark propagator:
\begin{equation}
{1 \over [\ti{p}-\ti{k}]^2 + m^2_D}\ =\ {Z_{pk} \over \ti{p}\,\ti{k}} \sum_{n=0}^\infty
\Big( Z_{pk} \Big)^n C_n \big(\hat{p} \cdot \hat{k} \big)\ ,
\label{eq:HS_denom}
\end{equation}
where explicitly,
\begin{equation}
Z_{pk}\ \equiv\ {1 \over 2 \ti{p}\,\ti{k}} \Big( \ti{p}\!\;^2 + \ti{k}^2 + m^2_D -
\sqrt{ (\ti{p}\!\;^2 + \ti{k}^2 +m^2_D)^2 - 4\ti{p}\!\;^2 \ti{k}^2 } \Big)\ ,
\label{eq:ZOFF}
\end{equation}
and we sometimes find it convenient to work in terms of the dimensionful object
$Z \equiv Z_{pk} \big/ \ti{p}\,\ti{k}$. In Eq.~(\ref{eq:HS_denom}), the $C_n$
are {\it Gegenbauer polynomials} with the normalization and orthogonality
properties described in App.~\ref{sec:appB}, and $\hat{p}$ is a unit
vector in Euclidean space in the direction of $\ti{p}_\mu$. We can exploit these
properties in App.~\ref{sec:appB} to perform the necessary angular integrations by first rendering the
numerator of Eq.~(\ref{eq:F1inner}) in terms of a linear combination of the Gegenbauer
polynomials
\begin{align}
\label{eq:pk}
(\ti{p}\cdot\ti{k})\ &=\ {\ti{p}\,\ti{k} \over 2}\, C_1(\hat{p} \cdot \hat{k})\ , \\
(\ti{p}\cdot\ti{k})^2\ &=\ {1 \over 4}\, \ti{p}\!\;^2\, \ti{k}^2\, \Big( C_2(\hat{p}\cdot\hat{k})
+ C_0(\hat{p}\cdot\hat{k}) \Big)\ .
\end{align}

Inserting everything into Eq.~(\ref{eq:F1inner}) and using Eq.~(\ref{eq:radint}) then results in 
\begin{align}
F_1(0)\ &=\ {g^2 \Lambda^4 \over (2\pi)^4}\ \int {d\ti{k}^2 \over 2}\
{\ti{k}^2 Z \over (\ti{k}^2+m^2)^2\, (\ti{k}^2+\Lambda^2)^2 } \int d\Omega_{\hat{k}}\
\Big( \sum_{n=0}^\infty \big( \ti{p}\,\ti{k}\, Z \big)^n C_n(\hat{p}\cdot\hat{k}) \Big) \nonumber\\
&\times\ \left( -{\ti{k}^2 \over 2} \big( C_2(\hat{p}\cdot\hat{k}) + C_0(\hat{p}\cdot\hat{k}) \big)
+ {m M \over \ti{p}\!\;^2}\ti{p}\,\ti{k}\, C_1(\hat{p}\cdot\hat{k}) + (m^2+\ti{k}^2)\, C_0(\hat{p}\cdot\hat{k})
\right)\ ;
\label{eq:F1_OFFS}
\end{align}
and we may use Eq.~(\ref{eq:orth}) to evaluate the angular integral $\int \Omega_{\hat{k}}$. Before
doing so, however, it is imperative to note that Eq.~(\ref{eq:F1_OFFS}) is defined in general for
spacelike $4$-momenta (including the external nucleon $4$-momentum $\ti{p}^2 \ge 0$). It is therefore
necessary to perform an analytic continuation of the proton momentum into the timelike region where
it is explicitly on-shell and thus physical: $\ti{p}\!\;^2 = -M^2$. By merit of our requirement that
$m_D \ge M$, the integration contour $\ti{k}^2 \in [0, \infty)$ remains unmenaced by branch points or
singularities, and the nucleon momentum may be straightforwardly continued to $\ti{p} \rightarrow iM$.
Doing so after evaluating the angular integrals, we finally obtain
\begin{equation}
F_1(0)\ =\ \left( {g \Lambda^2 \over 4\pi} \right)^2\ \int d\ti{k}^2\
{\ti{k}^2 \overline{Z} \over (\ti{k}^2+m^2)^2\, (\ti{k}^2+\Lambda^2)^2 }\
\left( {\ti{k}^2 \over 2} + {M^2 \over 2} \big( \ti{k}^2 \overline{Z} \big)^2
+ m M \ti{k}^2 \overline{Z} + m^2 \right)\ ,
\label{eq:EM-hyp}
\end{equation}
in which $\overline{Z}$ represents the analytic continuation of the rational function
$Z$ of Eq.~(\ref{eq:ZOFF}), given explicitly by
\begin{equation}
\overline{Z}\ =\ -{1 \over 2 M^2 \ti{k}^2}\ \left( \ti{k}^2 + \delta^2 -
\sqrt{ (\ti{k}^2 + \delta^2)^2 + 4 M^2 \ti{k}^2 } \right)\ ,
\label{eq:ZOS}
\end{equation}
having defined the shorthand $\delta^2 \equiv m_D^2 - M^2$.

It is notable also that the expression given in Eq.~(\ref{eq:EM-hyp}) constitutes an important check of the
hyperspherical formalism which we use in Sec.~\ref{sec:a0} below for $a_0$, and one may straightforwardly
verify that it yields $F_1(0)=1$ for the parameters of Table~\ref{tab:fit}. From it, we may at last extract
the Euclidean density function $\overline{f}_1(\ti{k}^2)$ for the proton's quark-level charge through direct
matching with Eq.~(\ref{eq:F1mom}),
\begin{equation}
\overline{f}_1(\ti{k}^2)\ =\ \left( {g \Lambda^2 \over 4\pi} \right)^2\
{\ti{k}^2 \overline{Z} \over (\ti{k}^2+m^2)^2\, (\ti{k}^2+\Lambda^2)^2 }\
\left( {\ti{k}^2 \over 2} + {M^2 \over 2} \big( \ti{k}^2 \overline{Z} \big)^2
+ m M \ti{k}^2 \overline{Z} + m^2 \right)\ ;
\label{eq:EM-f1}
\end{equation}
we plot this EDF in Fig.~\ref{fig:F1a0} alongside the analoguous quantity for the axial-singlet
charge $\overline{a}_0(\ti{k}^2)$ derived in Sec.~\ref{sec:a0} below.

Having determined the quark-level EDF for the proton's electric charge in Eq.~(\ref{eq:EM-f1}), we
may use this result to evaluate higher moments of the charge distribution given in
Eq.~(\ref{eq:f1k2_def}):
\begin{equation}
M^1_{\bar{f}_1}\ =\ 0.1985\ \mathrm{GeV}^2\ .
\label{eq:f1k2}
\end{equation}
In this case, this value corresponds roughly to the center of the peak of the heavy-solid line
in Fig.~\ref{fig:F1a0}; more directly, we also plot the integrand over $\ti{k}$ for the moment
$M^1_{\bar{f}_1}$ as the thin-solid line, multiplied by a factor of $2$ for ease of comparison.
%
%
%
\subsection{Quark helicity}
\label{sec:a0}
While the formalism in Sec.~\ref{sec:a0c} above was sufficient to determine the bare quark helicity contribution
to the proton axial-singlet charge $a_0$, we must ultimately interface our quark-diquark framework with the
results of BSE analyses to estimate the gluon dressing effect as mentioned above. In this case, the BSE
calculations we aim to incorporate are $\ti{k}$-dependent vertex factors as noted in Sec.~\ref{sec:EDF}, and
thus we require an axial charge momentum distribution along the lines of Eq.~(\ref{eq:EM-f1}) to evaluate
Eq.~(\ref{eq:aprime}).
\begin{figure}
\includegraphics[scale=0.55]{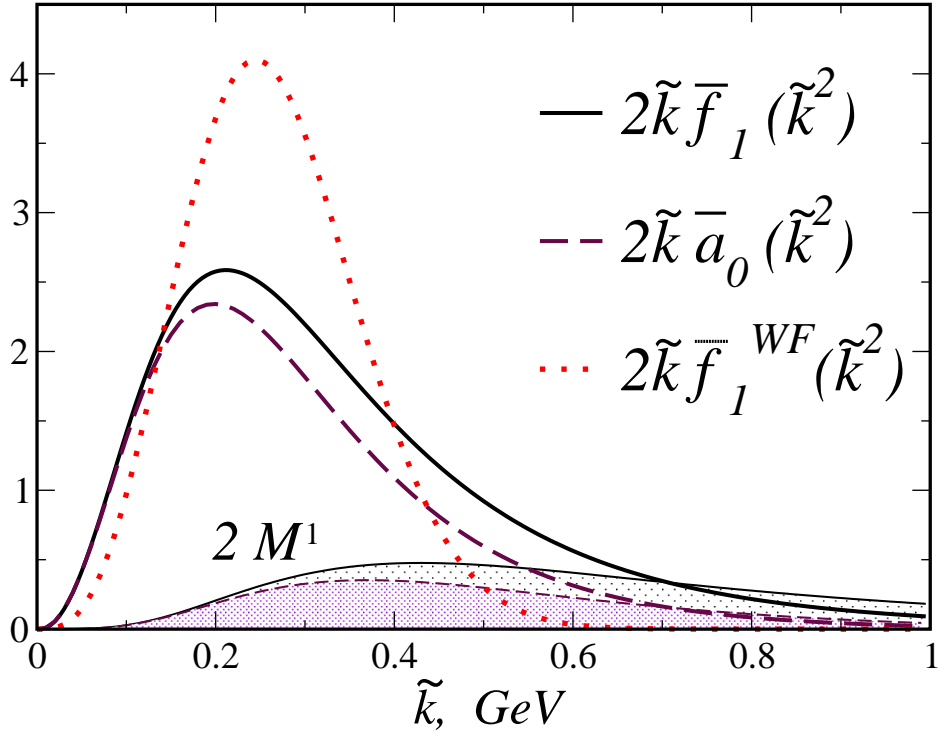}
\vspace*{-0.5cm}
\caption{ (Color online)
A comparison of EDFs for the proton's charge $2\ti{k}\, \overline{f}_1(\ti{k}^2)$ [Eq.~(\ref{eq:EM-f1}), black-solid]
and axial-singlet charge $2\ti{k}\, \overline{a}_0(\ti{k}^2)$ [Eq.~(\ref{eq:HS_ker}), maroon-dashed]
carried by the struck quark in the scalar diquark ECQM as functions of its Euclidean momentum
$\ti{k}$; for illustration, we contrast these with the result of using the Gaussian wave function,
$2\ti{k}\, \overline{f}^{\mathrm{WF}}_1(\ti{k}^2)$ from Eq.~(\ref{eq:f1WF}) [red-dotted].
The thin lines and associated shaded regions at bottom correspond to the integrands of these
distributions' first  moments in $\ti{k}^2$, \IE~$M^1 \sim \lan \ti{k}^2 \ran$ of Eqs.~(\ref{eq:f1k2_def})
and (\ref{eq:a0k2_def}). Note that these latter moments have been rescaled by a factor of $2$ for
comparison.
}
\label{fig:F1a0}
\end{figure}

Hence, analogously to the calculation in Sec.~\ref{sec:F1}, we now proceed by contracting both sides of Eq.~(\ref{eq:a0_amp})
with the nucleon spin $4$-vector $\ti{S}_\mu$ to obtain
\begin{align}
2M a_0\ &=\ -{g^2 \Lambda^4 \over (2\pi)^4}\ \int d^4 \ti{k}\
{\ti{S}_\mu\ \overline{u}(\ti{p}\!\;)
\left( { -i\ti{\ksl} + m } \right)\, \gamma_\mu \gamma_5\,
\left( { -i\ti{\ksl} + m } \right) u(\ti{p}\!\;)
\over (\ti{k}^2+m^2)^2\, (\ti{k}^2+\Lambda^2)^2\, \Big( [\ti{p}-\ti{k}]^2+m^2_D \Big) } \nonumber\\
&=\ -{g^2 \Lambda^4 \over (2\pi)^4}\ \int d^4 \ti{k}\
{2M\, \big(\, 2(\ti{S} \cdot \ti{k})^2 + (m^2 - \ti{k}^2)\ti{S}^2\, \big)\ -\ 4m (\ti{p} \cdot \ti{k})
\over
(\ti{k}^2+m^2)^2\, (\ti{k}^2+\Lambda^2)^2\, \Big( [\ti{p}-\ti{k}]^2+m^2_D \Big) }\ ,
\label{eq:HS_a0}
\end{align}
and here we require an additional inner product:
\begin{equation}
(\ti{S}\cdot\ti{k})^2\ =\ {1 \over 4}\,\ti{S}^2 \ti{k}^2\, \Big( C_2(\hat{S}\cdot\hat{k})
+ C_0(\hat{S}\cdot\hat{k}) \Big)\ .
\end{equation}
Using this and Eq.~(\ref{eq:pk}) to re-write the inner products of Eq.~(\ref{eq:HS_a0}) above, we 
incorporate the polynomial expansion for $\big(\,[\ti{p}-\ti{k}]^2+m^2_D\,\big)^{-1}$; here this leads to

\begin{align}
\label{eq:HS_inn}
a_0\ &=\ {g^2 \Lambda^4 \over (2\pi)^4}\ \int {d\ti{k}^2 \over 2}\
{\ti{k}^2 Z \over (\ti{k}^2+m^2)^2\, (\ti{k}^2+\Lambda^2)^2 } \int d\Omega_{\hat{k}}\
\Big( \sum_{n=0}^\infty \big( \ti{p}\,\ti{k}\, Z \big)^n C_n(\hat{p}\cdot\hat{k}) \Big) \\
&\ \ \ \ \ \times\ \Big( {\ti{k}^2 \over 2} \big( C_2(\hat{S}\cdot\hat{k}) + C_0(\hat{S}\cdot\hat{k}) \big)
-{m \over M}\,\ti{p}\,\ti{k}\, C_1(\hat{p}\cdot\hat{k}) + (m^2-\ti{k}^2)\, C_0(\hat{S}\cdot\hat{k})
\Big) \nonumber\\
&=\ \left( {g \Lambda^2 \over 4\pi} \right)^2\ \int d\ti{k}^2\
{\ti{k}^2 Z \over (\ti{k}^2+m^2)^2\, (\ti{k}^2+\Lambda^2)^2 }\ \nonumber\\
&\ \ \ \ \ \times\ \Big( {\ti{k}^2 \over 2} \big( \ti{p}\,\ti{k}\, Z \big)^2  {C_2(\hat{S}\cdot\hat{p}) \over 3}
-{m \over M}\ \ti{p}\,\ti{k}\, \big( \ti{p}\,\ti{k}\, Z \big) {C_1(\hat{p}\cdot\hat{p}) \over 2}
+ (m^2-{\ti{k}^2 \over 2})\, C_0(\hat{S}\cdot\hat{p})
\Big)\ .
\label{eq:HS_Geg}
\end{align}
As before, we analytically extend $\ti{p}$ into the timelike region where it is
on-shell, leading to 
\begin{align}
a_0\ &=\ \left( {g \Lambda^2 \over 4\pi} \right)^2\ \int d\ti{k}^2\
{\ti{k}^2 \overline{Z} \over (\ti{k}^2+m^2)^2\, (\ti{k}^2+\Lambda^2)^2 }\
\left( -{\ti{k}^2 \over 2} + {M^2 \over 6} \big( \ti{k}^2 \overline{Z} \big)^2 + m M \ti{k}^2 \overline{Z} + m^2 \right)\ ,
\label{eq:HS_fin}
\end{align}
and $\overline{Z}$ is again given by the expression in Eq.~(\ref{eq:ZOS}).
Lastly, we deduce the EDF appearing in Eq.~(\ref{eq:a0mom}) [and Eq.~(\ref{eq:aprime})] from Eq.~(\ref{eq:HS_fin})
by simple matching, as had been done for $\bar{f}_1(\ti{k}^2)$:
\begin{align}
\overline{a}_0(\ti{k}^2)\ =\ \left( {g \Lambda^2 \over 4\pi} \right)^2\
{\ti{k}^2 \overline{Z} \over (\ti{k}^2+m^2)^2\, (\ti{k}^2+\Lambda^2)^2 }\
\left( -{\ti{k}^2 \over 2} + {M^2 \over 6} \big( \ti{k}^2 \overline{Z} \big)^2
+ m M \ti{k}^2 \overline{Z} + m^2 \right)\ ;
\label{eq:HS_ker}
\end{align}
in summary, we emphasize that to obtain Eqs.~(\ref{eq:HS_a0})--(\ref{eq:HS_ker}) we have contracted both sides of the first
equation with $\ti{S}_\mu$ and expanded the diquark propagator {\it \`a la} Eq.~(\ref{eq:HS_denom}).

With these expressions, one may proceed to compute the bare quark contribution to the proton
spin using the set of parameters determined from fits to the proton electromagnetic form factors,
given in Table~\ref{tab:fit}. Using these values in the conventional formalism of Sec.~\ref{sec:a0c}
that led to Eq.~(\ref{eq:a0_cov2}), we found $a_0 = 0.784$ --- a value which may also be recovered
from the hyperspherical formalism as given by Eq.~(\ref{eq:HS_fin}). Incidentally, this figure is
in accord with the moment of the scalar diquark contribution to the quark helicity PDF obtained
in a typical light-front quark model (see Eqs.~(61) and (62) of Ref.~\cite{Cloet:2012cy}):
\begin{equation}
\Delta q_s\ =\ {1 \over 3}\, \left( 2\, \Delta u\ -\ \Delta d \right)\ \approx\ 0.75\ ;
\label{eq:LF-heli}
\end{equation}
this latter expression assumed an $\mathbf{SU(2) \otimes SU(2)}$ structure for the proton's
spin-flavor wave function.

We point out as well that the axial-singlet
EDF $\overline{a}_0(\ti{k}^2)$ given by Eq.~(\ref{eq:HS_ker}) is not restricted to be positive-definite, unlike
the analogous electromagnetic charge EDF $\overline{f}_1(\ti{k}^2)$ of Eq.~(\ref{eq:EM-f1}), which is related
to the zeroth moments of traditional probabilistic quark density functions. In fact, for certain parameter
combinations, $\overline{a}_0(\ti{k}^2)$ may experience substantial negative downturns at larger spacelike
quark momenta, $\ti{k} \ge 1$ GeV. However, for the set of fitting parameters that best describes proton
form factor data, this effect is not evident, and the axial-singlet EDF $\overline{a}_0(\ti{k})$ is
instead dominated by a soft peak centered roughly at $\ti{k} \lesssim 0.2$ GeV, as shown in Fig.~\ref{fig:F1a0}
as the maroon-dashed line.
\begin{figure}
\includegraphics[scale=0.46]{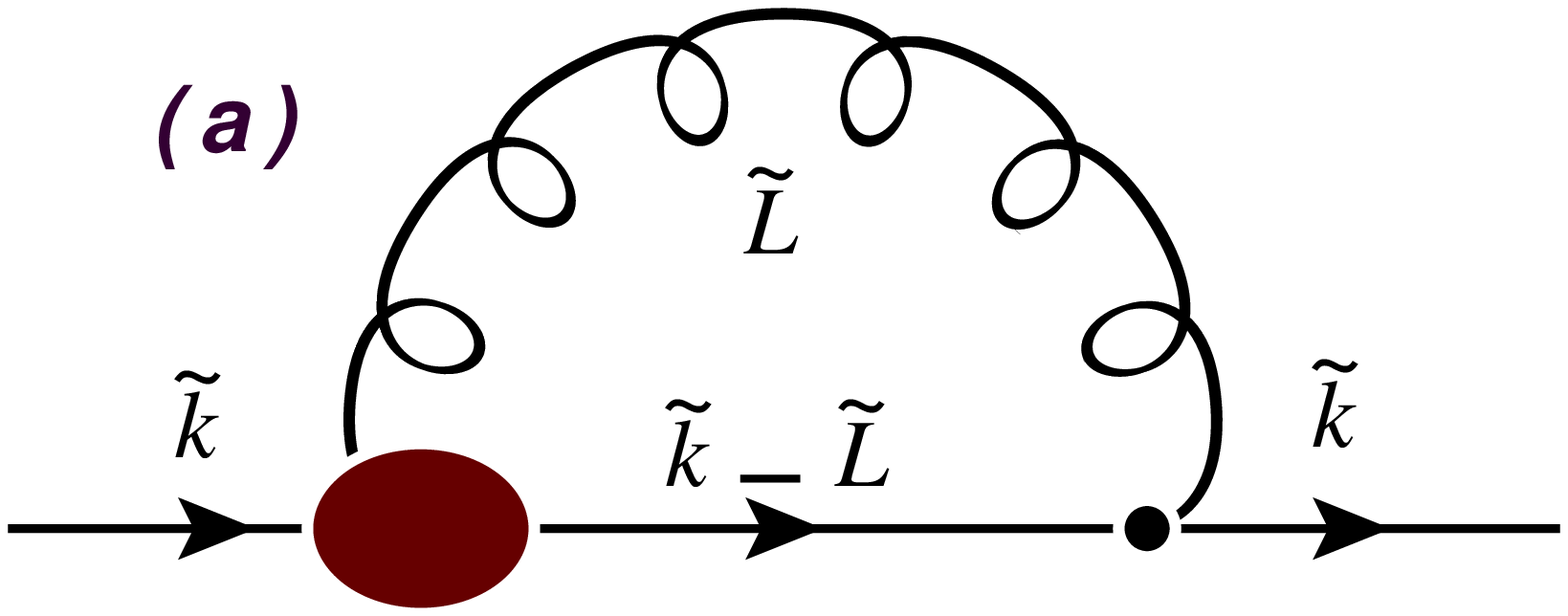}\ \ \ \
\raisebox{-0.9cm}{\includegraphics[scale=0.46]{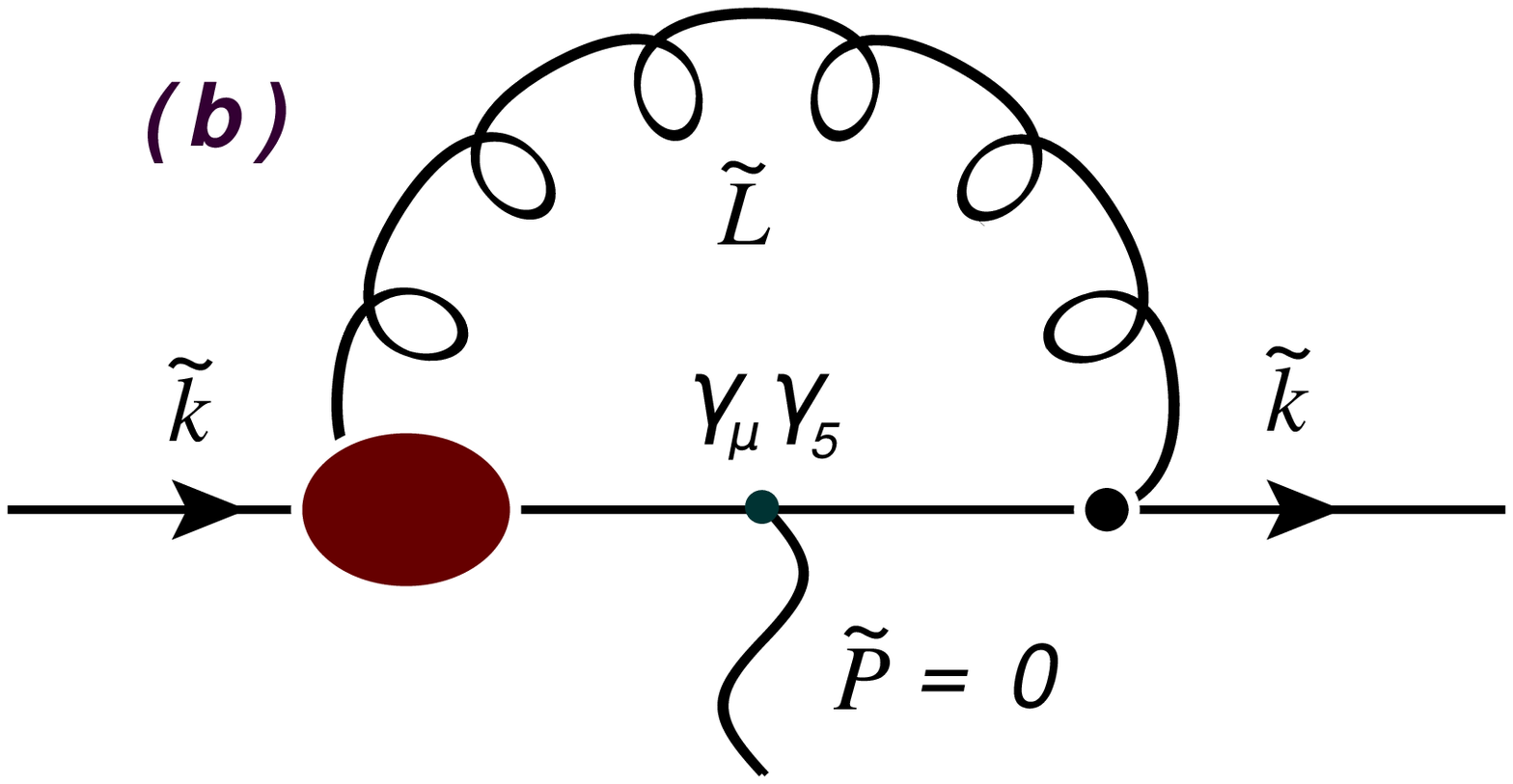}}
\vspace*{-0.2cm}
\caption{ (Color online)
({\bf a}) The diagram leading to the DSE for a quark of momentum
$\ti{k}$ dressed by a nonperturbative gluon carrying loop momentum $\ti{L}$.
({\bf b}) The corresponding diagram for the quark axial-vector vertex BSE, responsible
for the $\ti{k}$-dependent gluonic dressing correction to the axial charge of an
individual quark.
}
\label{fig:DSE}
\end{figure}
Owing mainly to the similarity of the explicit $\ti{k}^2$ dependence appearing in
Eqs.~(\ref{eq:EM-f1}) and (\ref{eq:HS_ker}), the shapes of these distributions closely track each
other, with $f_1(\ti{k}^2) \approx \overline{a}_0(\ti{k}^2)$, particularly for $\ti{k}^2 \ll m^2$.
Ultimately, we interpret this behavior as following from the common origin of both expressions
in the diagrams of Fig.~\ref{fig:FF}, which at $\ti{q}=0$ differ only by the appearance of $\gamma_5$.

Moreover, for the higher $\sim\!\lan \ti{k}^2 \ran$ moment of the axial-singlet EDF, we obtain the value
\begin{equation}
M^1_{\bar{a}_0}\ =\ 0.08125\ \mathrm{GeV}^2\ ,
\label{eq:a0k2}
\end{equation}
implying the proton's distribution of axial-singlet charge is relatively softer than the
charge distribution [Eq.~(\ref{eq:f1k2})] in the bare model.
%
%
\section{Gluon Dressing Effect}
\label{sec:gluon}
We now incorporate numerical estimates of the effect of dressing the quark-axial current vertex
with gluon exchange, which in principle may be determined from DSE-BSE analyses. Here, the relevant
diagrams are displayed in Fig.~\ref{fig:DSE}, wherein panel ({\it\bf a}) illustrates the dressed propagator
responsible for QCD's quark DSE, while panel ({\it\bf b}) demonstrates the realization of the BSE for the
quark-level coupling of the axial-vector current dressed by soft gluon exchange. Naturally, the infrared
momenta at which this effect is of interest demands the use of nonperturbative methods, and the standard
procedure requires a prescription-dependent truncation of the quark-gluon vertex (shown as the blobs
in Fig.~\ref{fig:DSE}). 

In the context of BSE analyses \cite{Maris:1997hd,Bhagwat:2002tx,Bhagwat:2007ha,Chang:2012cc}, the
dressed axial-vector vertex is represented by the structure $\Gamma^{fg}_{5\mu}(\ti{K};\ti{P})$, which is
understood to connect an incoming quark of flavor $g$ and momentum $\ti{K}_- = \ti{K} - (1-\eta)\ti{P}$ to
an outgoing quark of flavor $f$ and momentum $\ti{K}_+ = \ti{K} + \eta \ti{P}$; here $\ti{P}$ and $\ti{K}$
represent the total and relative momentum of the quark pair, and $\eta$ is a dimensionless parameter upon
which calculations cannot depend. Thus, for our purposes, we require the case $\ti{P}=0$, such that
$\ti{K}_+ = \ti{K}_- = \ti{K} \equiv \ti{k}$, and we take the diagonal isospin-independent vertex $f=g$,
as described in Ref.~\cite{Chang:2012cc}.
Then the structure of the quark-axial vector vertex of relevance here is simply
\begin{equation}
\overline{u}(\ti{k})\ \Gamma_{5\mu}(\ti{k};0)\ u(\ti{k})\ =\
\overline{u}(\ti{k})\ \gamma_5 \left[ \gamma_\mu F_R (\ti{k};0) + \dots \right]\ u(\ti{k})\ ,
\label{eq:AV_vert}
\end{equation}
and the ellipsis in Eq.~(\ref{eq:AV_vert}) above represents additional contributions to the
vertex that do not contribute in the present analysis. We therefore make the identification
$f_g(\ti{k}) \equiv F_R(\ti{k};0)$ mentioned in Sec.~\ref{sec:EDF}, and directly insert the numerical
results reported in Ref.~\cite{Chang:2012cc} to smear the bare model axial charge as in Eq.~(\ref{eq:aprime}).

\begin{figure}
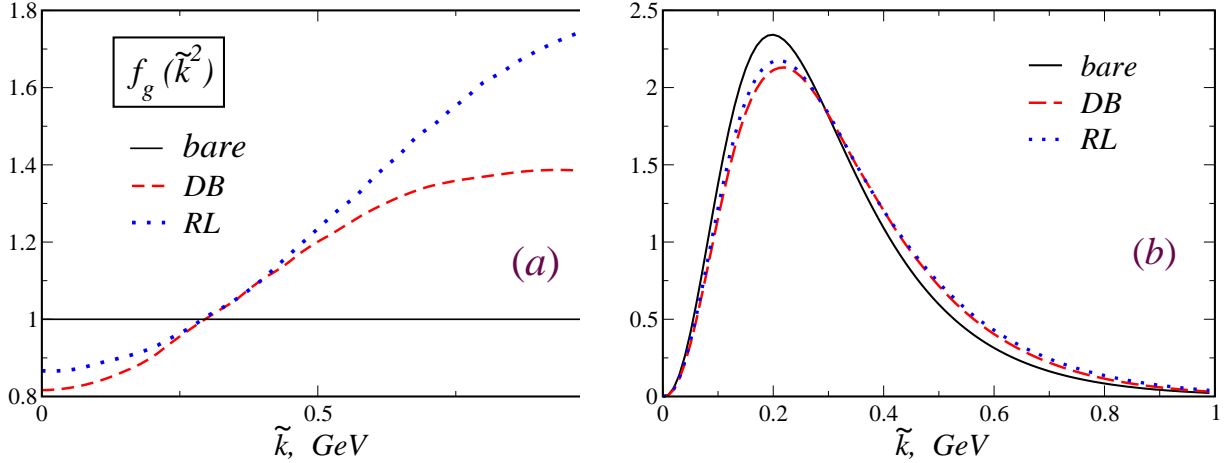

\includegraphics[scale=0.34]{F5a.eps} \ \
\includegraphics[scale=0.34]{F5b.eps}
\caption{ (Color online)
({\bf a}) The gluon dressing function $f_g(\ti{k}^2)$ under several different scenarios:
the perturbative limit $f_g(\ti{k}^2)=1$ (solid black), and $f^{\mathrm{DB}}_g(\ti{k}^2)$ (red-dashed)
and $f^{\mathrm{RL}}_g(\ti{k}^2)$ (blue-dotted).
({\bf b}) A plot of the integrand of Eq.~(\ref{eq:aprime}) $2\ti{k}\, \overline{a}_0(\ti{k}^2)\, f_g(\ti{k}^2)$ for several
choices of the gluon dressing function: $f_g(\ti{k}^2)=1$ (``{\it bare},'' shown in solid black), as well as
the result of an improved dynamical chrial symmetry-breaking kernel in the BSE
$f^{\mathrm{DB}}_g(\ti{k}^2)$ (``DB,'' red-dashed line), and the rainbow-ladder truncation method
$f^{\mathrm{RL}}_g(\ti{k}^2)$ (``RL,'' blue-dotted curve) of Ref.~\cite{Chang:2012cc,Maris:1997hd}.
}
\label{fig:FR}
\end{figure}

The behavior of $f_g(\ti{k})$ depends crucially on the truncation scheme used to obtain the effective
quark-gluon vertices in the panels of Fig.~\ref{fig:DSE}. To get a sense for this source of
prescription dependence, we compute the correction following from both schemes treated in
Ref.~\cite{Chang:2012cc} --- the rainbow-ladder scheme (RL), and an ansatz based on a specific
realization of dynamical symmetry breaking (DB), which we take numerically from Fig.~1 of
Ref.~\cite{Chang:2012cc}. Referring to these as $f^{\mathrm{RL}}_g(\ti{k})$ (blue-dotted) and
$f^{\mathrm{DB}}_g(\ti{k})$ (red-dashed), we plot both dressing functions against $\ti{k}$ in
Fig.~\ref{fig:FR}({\bf a}). Plainly, both truncation shemes predict a suppression of the quark's
axial charge for the lowest infrared momenta $\ti{k} \lesssim 0.3$ GeV, but substantial enhancements
beyond --- particularly for the RL prescription, which overhangs the DB scheme by $\sim\! 25\%$ for
$\ti{k} \sim 1$ GeV.
Having determined the axial EDF of Eq.~(\ref{eq:HS_ker}) we may fold these extractions for the gluon
dressing function into Eq.~(\ref{eq:aprime}) to determine the overall effect, plotting the integrands
responsible for $a'_0$, $2\ti{k}\, f^{\mathrm{ RL}}_g(\ti{k}^2)\, \overline{a}_0({\ti{k}^2})$ (blue-dotted)
and $2\ti{k}\, f^{\mathrm{DB}}_g(\ti{k}^2)\, \overline{a}_0({\ti{k}^2})$ (red-dashed), in
Fig.~\ref{fig:FR}({\bf b}) alongside the bare or ``undressed'' scenario, $f_g(\ti{k}^2) = 1$ (black-solid).
From this, we find the net correction to the quark helicity contribution from gluon dressing to be 
\begin{align}
\label{eq:DBnum}
\left( {a'_0 \over a_0} \right)\ - 1\ &=\ -0.04\%\ \ \ \  (\mathrm{DB\ scheme})\ , \\
                        &=\ +2.98\%\ \ \ \  (\mathrm{RL\ scheme})\ .
\label{eq:RLnum}
\end{align}
The magnitude of the effect from gluon dressing is therefore quite small, and in the
present analysis, actually consistent with zero in the sense that depending upon the
choice of truncation scheme, one may obtain a modest enhancement (RL) or tiny suppression
(DB) of the proton's total quark helicity. The smallness of the effect can be understood
from the momentum dependence shown in Fig.~\ref{fig:FR}({\bf b}), in which the interplay of the
shapes of $f_g(\ti{k}^2)$ and $\overline{a}_0(\ti{k}^2)$ are such that the axial-singlet charge
is slightly suppressed at low $\ti{k}$ and enhanced at higher $\ti{k}$. These two effects
largely cancel, however, in the integral over $\ti{k}$ involved in the computation of $a'_0$ according
to Eq.~(\ref{eq:aprime}), such that $a'_0 \approx a_0$, and we conclude the dressing effect in $a_0$
to be minimal.
%
%
%
%
\section{Conclusion}
\label{sec:conc}
%
In this paper we have proposed a model in Euclidean space formulated in terms of constituent quark
degrees of freedom. The essential products of the resulting ECQM technology are density functions of the
quark's Euclidean momentum (the EDFs) obtained from hyperspherical angular integrations of
$4$-dimensional amplitudes. The special value of these derived quantities is their ability to
recover nucleon charges through integrals over the internal momenta of their constituent quarks,
a fact that empowered us to couple them to predictions of other Euclidean analyses --- in this case,
BSEs.

Thus, having introduced this formalism, we tested it preliminarily by computing both
the nucleon's quark charge density, as well as its axial-singlet charge. For the
latter, this test assumed the form of an assessment of the impact of BSE calculations for the dressed
quark axial-vector vertex. There are of course various sources of model dependence on the
side of both our ECQM for the nucleon-quark interaction and of the BSE analyses. Despite
these sources of model-dependence, we find the effect of the gluon dressing to be small
--- at most a several percent correction to the total quark helicity in the bare ECQM.

Naturally, the analysis presented here is essentially exploratory, and if
anything, suggests the need for further refinements. For instance, the scalar diquark
picture alone cannot realistically approximate the nucleon's full spin structure as evidenced by
the large value we obtain for the bare axial-singlet charge ($a_0 = 0.784$); a
fuller calculation would therefore involve spin-$1$ diquark exchanges, which in general are
necessary to obtain an authentic flavor decompostion of the nucleon helicity. At the same time, it
is reasonable to expect that the qualitative {\it shape} obtained for $\overline{a}_0(\ti{k})$ shown in
Fig.~\ref{fig:FR} for the present scalar diquark ECQM would hold also for amplitudes involving
spin-$1$ exchanges, so that the essential details of such a calculation would resemble
our presentation here. That being the case, our ultimate conclusion is unlikely to change: models
in which bare constituent quarks carry the great predominance of the total quark helicity are on
robust footing.

Similarly, it should be noted that other possible considerations have not been treated systematically,
including the momentum dependence of the constituent quark's dynamical mass, the implementation
of which would require a self-consistent scheme not typical of the fitted constituent quark model
presented here. Such issues, as well as continued improvements to the Euclidean hyperspherical
formalism and BSEs for the axial-vertex dressing functions will be of enormous value in extending
the current state-of-the-art regarding quark helicity, the nucleon spin problem, and Euclidean
modeling of nucleon structure.
%
%
%
\section{Acknowledgements}
We thank Ian Clo\"et, Javier Men\'endez, Brian Tiburzi, Andre Walker-Loud,
and Xilin Zhang for helpful exchanges. The work of TJH and GAM was supported
by the U.S.~Department of Energy Office of Science, Office of Basic Energy
Sciences program under Award Number DE-FG02-97ER-41014. The work of MA was
supported under NSF Grant No.~1516105.
%
%
\appendix
\section{Euclidean space conventions}
\label{sec:appA}
%
We proceed using the Minkowski $\leftrightarrow$ Euclidean transcription dictionary as outlined in, \EG~Refs.~\cite{Roberts:1994dr,Cloet:2013jya}, wherein
$4$-momenta and Dirac matrices transform according to
\begin{align}
k^0 &= ik_4\ , \hspace{1.5cm} k^j = -k_j\ , \nonumber\\
\gamma^0 &= \gamma_4\ , \hspace{1.65cm} \gamma^j = i \gamma_j\ ; \hspace{1.5cm} j \in \{1,2,3\}\ .
\label{eq:Eucl-momenta}
\end{align}
The Dirac algebra in this setting is then specified by
\begin{equation}
\Big\{ \gamma_\mu , \gamma_\nu \Big\}\ =\ 2\, \delta_{\mu\nu}\ ,
\end{equation}
such that the Euclidean inner product for any two $4$-vectors
$\ti{a}_\mu\,, \ti{b}_\mu$ is 
\begin{equation} 
\ti{a} \cdot \ti{b}\ \equiv\ \sum_\mu\ \ti{a}_\mu \ti{b}_\mu\ =\ \ti{a}_1\, \ti{b}_1 + \dots + \ti{a}_4\, \ti{b}_4\ ,
\end{equation} 
and, by extension,
\begin{equation}
\ti{\psl}\ \equiv\ \gamma_1\,\ti{p}_1 + \dots + \gamma_4\,\ti{p}_4\ .
\end{equation}
We also note the definition
\begin{equation}
\gamma_5\ =\ - \gamma_1\, \gamma_2\, \gamma_3\, \gamma_4\ .
\end{equation}
We may give explicit expressions for the Euclidean Dirac spinors, which we obtain
following the conventional Wick rotation as
\begin{align}
u_\lambda(p)\ =\ \sqrt{M + p^0}\
\left( \begin{array}{c}
\chi_\lambda \\
{{\mathbf \sigma} \cdot {\bf p} \over M + p^0}\ \chi_\lambda
\end{array} \right) \hspace*{0.5cm}
&\rightarrow \hspace*{0.5cm}
u_\lambda(\ti{p}\!\;)\ =\ \sqrt{M + i\ti{p}_4}\
\left( \begin{array}{c}
\chi_\lambda \\
{-{\mathbf \sigma} \cdot {\bf \ti{p}} \over M + i\ti{p}_4}\ \chi_\lambda
\end{array} \right)\ ,
\label{eq:spinors}
\end{align}
where the helicity states $\chi_{[\lambda = \uparrow\downarrow]} = \left( \begin{array}{c} 1 \\ 0 \end{array} \right),\ 
\left( \begin{array}{c} 0 \\ 1 \end{array} \right)$ are proportional to the standard
eigenvectors of $\sigma_3$. These spinors are endowed with the typical normalization,
\begin{equation}
\overline{u}\, u\ =\ 2 M , \hspace*{1cm}
\overline{u}(\ti{p}\!\;)\, \gamma_\mu\, u(\ti{p}\!\;) = 2i\, \ti{p}_\mu\ ,
\end{equation}
and obey the Dirac Equation
\begin{equation}
\overline{u}(\ti{p}\!\;^\prime)(i\ti{\psl}' + M) = (i\ti{\psl} + M)u(\ti{p}\!\;) = 0\ .
\label{eq:EuDir}
\end{equation}
Moreover, in Euclidean space, the Gordon Identity assumes the slightly altered form
\begin{equation}
\overline{u}(\ti{p}\!\;^\prime)\, \gamma_\mu\, u(\ti{p}\!\;)\ =\ {1 \over 2 M}\ \overline{u}(\ti{p}\!\;^\prime)
\left\{ -i \ti{P}_\mu\ +\ \sigma_{\mu\nu} \ti{q}_\nu \right\} u(\ti{p}\!\;)\ ,
\label{eq:Gordon}
\end{equation}
where we have defined $\ti{P}_\mu \equiv \ti{p}\!\;^\prime_\mu + \ti{p}_\mu$ and $\sigma_{\mu\nu} \equiv (i/2)[\gamma_\mu,\gamma_\nu]$.
By similar logic, we obtain the general form for the extended electromagnetic vertex of the proton,
\begin{equation}
\overline{u}(\ti{p}\!\;^\prime)\, \Gamma_\mu \big( \ti{p}\!\;^\prime, \ti{p}\!\; \big)\, u(\ti{p}\!\;)\ =\ \overline{u}(\ti{p}\!\;^\prime)
\left\{ F_1(\ti{q}\!\;^2)\, \gamma_\mu\ +\ F_2(\ti{q}\!\;^2)\, \sigma_{\mu\nu} {\ti{q}_\nu \over 2M} \right\}
u(\ti{p}\!\;)\ .
\label{eq:EMV}
\end{equation}
%
\section{Hyperspherical formalism}
\label{sec:appB}
In the hyperspherical formalism \cite{Blum:2002ii,Levine:1974xh,Roskies:1990ki,Rosner:1967zz}, numerator
algebra leads to covariant expressions involving inner products which we represent in terms of the Gegenbauer
polynomials, of which only the lowest are relevant for the present analysis:
\begin{align}
C_0(x)\ &=\ 1\ , \hspace*{1.5cm}\ C_1(x)\ =\ 2x\ , \nonumber\\ 
C_2(x)\ &=\ 4x^2 - 1\ .
\end{align}
Hyperspherical integrals may be separated into radial and angular parts according to
\begin{equation}
\int\, d^d \ti{k}\ =\ \int\, d\ti{k}\ \ti{k}^{d-1}\ \int\, d\Omega^{(d)}_{\hat{k}}\ ,
\label{eq:radint}
\end{equation}
and we of course take $d=4$ in the integrations over
$d\Omega_{\hat{k}} \equiv d\Omega^{(4)}_{\hat{k}}\, \big(\!= \sin^2\psi \sin\theta\, d\phi\, d\theta\, d\psi \big)$
in Sec.~\ref{sec:HypS}; these can then be carried out in practice using well-known orthogonality properties:
\begin{equation}
\int d\Omega_{\hat{b}}\ C_m \big(\hat{a}\cdot\hat{b}\big)\ C_n \big(\hat{b}\cdot\hat{c}\big)\
=\ {2\pi^2 \delta_{mn} \over n+1}\ C_n \big( \hat{a}\cdot\hat{c} \big)\ .
\label{eq:orth}
\end{equation}
These relations can be determined from an appropriate choice of hyperspherical coordinates, with
a common selection \cite{Peskin:1995ev} being
\begin{align}
k_\mu\ =\ \sqrt{\,\ti{k}\!\;^2}\
\left( \begin{array}{c}
\sin\psi\,\sin\theta\,\cos \phi \\
\sin\psi\,\sin \theta\,\sin \phi \\
\sin\psi\,\cos\theta  \\
\cos\psi
\end{array} \right)\ .
\label{eq:coord}
\end{align}
%
%
%
%
%

%
\end{document}